\documentclass{article}

\usepackage{PRIMEarxiv}

\usepackage[utf8]{inputenc} 
\usepackage[T1]{fontenc}    
\usepackage{hyperref}       
\usepackage{url}            
\usepackage{booktabs}       
\usepackage{amsfonts}       
\usepackage{nicefrac}       
\usepackage{microtype}      
\usepackage{lipsum}
\usepackage{fancyhdr}       
\usepackage{graphicx}       

\usepackage[numbers]{natbib}
\usepackage{multirow}
\usepackage{subcaption}
\usepackage{amsmath}

\title{Privacy Preserving Data Imputation via Multi-party Computation for Medical Applications
}

\author{
  Julia Jentsch \thanks{These authors contributed equally to this work.} \\
  Medical Data Privacy and \\
  Privacy Preserving Machine Learning \\
  University of Tübingen \\
  Tübingen \\
  \texttt{julia.jentsch@student.uni-tuebingen.de} \\
   \And
  Ali Burak Ünal \footnotemark[1] \\
  Medical Data Privacy and \\
  Privacy Preserving Machine Learning \\
  University of Tübingen \\
  Tübingen \\
  \texttt{ali-burak.uenal@uni-tuebingen.de} \\
  \And
  Şeyma Selcan Mağara \\
  Medical Data Privacy and \\
  Privacy Preserving Machine Learning \\
  University of Tübingen \\
  Tübingen \\
  \texttt{seyma-selcan.magara@uni-tuebingen.de} \\
   \And
  Mete Akgün \\
  Medical Data Privacy and \\
  Privacy Preserving Machine Learning \\
  University of Tübingen \\
  Tübingen \\
  \texttt{mete.akguen@uni-tuebingen.de} \\
}

\begin{document}
\maketitle

\begin{abstract}
Handling missing data is crucial in machine learning, but many datasets contain gaps due to errors or non-response. Unlike traditional methods such as listwise deletion, which are simple but inadequate, the literature offers more sophisticated and effective methods, thereby improving sample size and accuracy. However, these methods require accessing the whole dataset, which contradicts the privacy regulations when the data is distributed among multiple sources. Especially in the medical and healthcare domain, such access reveals sensitive information about patients. This study addresses privacy-preserving imputation methods for sensitive data using secure multi-party computation, enabling secure computations without revealing any party's sensitive information. In this study, we realized the mean, median, regression, and kNN imputation methods in a privacy-preserving way.  We specifically target the medical and healthcare domains considering the significance of protection of the patient data, showcasing our methods on a diabetes dataset. Experiments on the diabetes dataset validated the correctness of our privacy-preserving imputation methods, yielding the largest error around $3 \times 10^{-3}$, closely matching plaintext methods. We also analyzed the scalability of our methods to varying numbers of samples, showing their applicability to real-world healthcare problems. Our analysis demonstrated that all our methods scale linearly with the number of samples. Except for kNN, the runtime of all our methods indicates that they can be utilized for large datasets.
\end{abstract}


\section{Introduction}
Among almost all domains, machine learning algorithms have shown significant progress and success in the healthcare and medical domains. However, these algorithms require the data to be complete and not to have a missing feature. Unfortunately, in real-world problems, but especially in medical and healthcare problems, the data has often some missing features due to utilizing multiple sources, sensor failures, and so on \cite{cismondi2013missing}. The simplest approach to handle such cases is to employ listwise deletion \cite{allison2009missing}. Although this approach is simple and computationally efficient, it tends to lead to biased estimates and reduces the sample sizes, which is especially problematic in medical and healthcare problems containing a limited number of samples \cite{Myers.2011}. Therefore, the researchers have proposed many sophisticated and effective methods to impute the missing data instead of discarding the whole sample completely.

Even though there exists a considerable amount of studies addressing how missing data can be handled, there are a limited number of approaches considering how such a case could be addressed when the data is distributed among multiple sources and their privacy has to be protected. Several approaches in the literature \cite{Omer.2017,Gursoy.2022,Du.2024} addressed the problem using homomorphic encryption as a privacy enhancing technique. However, due to the heavy encryption nature of homomorphic encryption, scalability and real-life deployment of such solutions are problematic. Differential privacy is also among the choices \cite{Das.30.06.2022}, but the noise introduced into the data may not be tolerable in many precision-requiring problems. As a promising alternative, secure multi-party computation (MPC) has drawn some attention as well, resulting in approaches addressing the missing data problem using MPC \cite{Jagannathan.2008}. However, they only utilized decision trees as an imputation method, leaving considerable room for improvement.


In this study, considering the balance between privacy and efficiency, we propose privacy-preserving data imputation methods using MPC and demonstrate their effectiveness and applicability to medical and healthcare problems on the diabetes dataset. Among the imputation methods proposed in the literature, we focus on mean and median imputations as starting approaches, and then move to more effective and sophisticated methods, namely regression and $k$-nearest neighbor (kNN) imputation methods. Considering the importance of scalable solutions in medical and healthcare domains, we also analyzed the execution time of our methods on varying datasets with varying numbers of samples which are sampled from the same dataset.

\section{Preliminary}

\subsection{Data Imputation Methods}
On top of the listwise deletion of samples with missing features, there are methods to impute missing values by replacing them with estimated ones. Imputation methods range from traditional mean imputation to more sophisticated techniques like regression imputation, allowing researchers to utilize all data, not just complete cases. To enable missing data imputation in an MPC framework, we selected $4$ widely used data imputation methods considering their effectiveness and computation feasibility in MPC.

\subsubsection{Mean Imputation}
One of the most common data imputation methods is the mean imputation where the missing feature of the samples is replaced with the mean of this feature of other samples \cite{kalton1982imputing}. The mean imputation involves calculating the mean of a numeric feature by adding all the available values and dividing by the number of non-missing values. The missing values are then replaced with this mean.


The mean imputation can be extended to categorical features by using the mode, or the most frequent value in the feature, as the mean of the feature. The simplicity of the mean imputation draws significant attention, but one must consider that the mean imputation can introduce bias and reduce variance \citep{Donders.2006}.

\subsubsection{Median Imputation}
As an alternative to the mean imputation, the literature offers a method called median imputation where the median feature value is assigned to the missing feature of the samples. The median imputation requires sorting the samples with non-missing data based on this feature and identifying the median sample. Then, the value of this sample's feature is used to impute the concerned feature of the samples with missing data. Similar to the mean imputation, even though the median imputation is simple and efficient, it can introduce bias and overlook the variance of the feature \citep{Zhang.2016}.

\subsubsection{Regression Imputation}
Regression imputation is a more advanced, multivariate method for addressing missing data. Instead of relying on a single statistic like the mean or median, this technique estimates the imputed value based on other features in the dataset \citep{Donders.2006}. The missing value is replaced with a predicted value derived from multiple linear regression using the non-missing data from other features. This method assumes a linear relationship between the features, which, if incorrect, could introduce bias. However, compared to mean or median imputation, regression imputation better preserves the original distribution of the data \citep{Jadhav.2019}. The multiple linear regression equation can be represented as follows:
\begin{equation}
    \mathbf{y} = \mathbf{X}\boldsymbol{\beta} + \boldsymbol{\varepsilon}
\end{equation}
where $\mathbf{y}$ is the vector of dependent variables, $\mathbf{X}$ is the design matrix, $\boldsymbol{\beta}$ is the vector of regression coefficients, and $\boldsymbol{\varepsilon}$ is the vector of residuals. The solution of this linear system can be rewritten by focusing on the regression coefficients using the Moore-Penrose inverse \citep{Sen.1990} as follows:
\begin{equation}
    \boldsymbol{\hat{\beta}} = (\mathbf{X}^\top \mathbf{X})^{-1} \mathbf{X}^\top \mathbf{y}
\end{equation}
Since the computation of inverse is a costly operation, we can rewrite the equation as follows:
\begin{equation}
    (\mathbf{X}^\top \mathbf{X}) \boldsymbol{\hat{\beta}} = \mathbf{X}^\top \mathbf{y}
\end{equation}
To obtain the linear coefficients $\boldsymbol{\hat{\beta}}$, even though QR decomposition is utilized, we will employ LU decomposition \citep{Golub.1988} considering its MPC-friendly operations and suboptimality of QR decomposition in MPC \cite{Liu.2020}. This allows us to obtain $Lz = X^{\top}y$ first where $z = U \boldsymbol{\hat{\beta}}$. The final step is to use backward substitution and obtain $\boldsymbol{\hat{\beta}}$.

\subsubsection{kNN Imputation}
kNN imputation, a relatively recent technique, is a type of hot-deck imputation method \citep{Beretta.2016}. It utilizes the $k$-nearest neighbor algorithm to estimate missing values by considering the values of the nearest neighbors. As a supervised learning algorithm, kNN is particularly useful for handling missing data. It identifies the variables most similar to the one with the missing value, known as the nearest neighbors, determined using a distance metric, typically Euclidean distance. However, to improve efficiency in a secure multi-party computation setting, the squared Euclidean distance is used considering that this does not affect the order of the samples while finding the $k$-nearest neighbors. The squared Euclidean distance is calculated as follows:

\begin{equation}
    \mathcal{E}_{dist}(\mathbf{x}, \mathbf{y}) = \sum_{i=1}^{d} (x_i - y_i)^2
    \label{eq:squared_euclidean_distance}
\end{equation}
where $\mathcal{E}_{dist}(x,y)$ is the squared Euclidean distance function, $\mathbf{x}$ and $\mathbf{y}$ are the feature vectors, $d$ is the dimension of the feature vectors. For numeric features, a weighted average of the $k$-nearest neighbors' values is used to estimate the missing value. For categorical features, the most frequent label among the neighbors is chosen. The parameter $k$, representing the number of neighbors considered, must be specified by the user, with a common heuristic being $\sqrt{N}$, the square root of the sample size \citep{Ispirova.2020}.

kNN imputation is known for its accuracy in preserving the original data structure and can handle non-linear relationships, unlike regression imputation which assumes linearity \citep{Liao.2014}. It is versatile, suitable for various types of features, making it effective for mixed-type data \citep{Kowarik.2016}. Despite its simplicity, kNN imputation is effective for many missing data problems.

However, kNN imputation also has challenges. Calculating distances can be computationally intensive, especially for large datasets, and the choice of \textit{k} can significantly impact imputation accuracy. Moreover, it is not robust to outliers, which can distort the imputed values of their nearest neighbors. Despite these drawbacks, kNN imputation remains a widely used and versatile method for filling in missing values, offering less bias compared to traditional techniques like mean imputation.

\subsection{Secure Multi-party Computation}
Among other privacy enhancing techniques, secure multi-party computation (MPC) provides privacy with an acceptable computational and communication overhead, without sacrificing the performance of the underlying method. The essence of MPC is to give a share or shares of the secret value to each computing party such that this share seems like a random value to the party. The only way to obtain the secret value is to bring all shares together, which is not permitted in a semi-honest adversarial setting. Even though these shares do not provide information about the secret value, the parties can still perform some operations on them without accessing the original value. There are several different MPC frameworks and approaches in the literature, providing different functionalities. Considering the function coverage and effectiveness, we decided to use CECILIA \cite{Unal.07.02.2022} to realize imputation methods in a privacy-preserving way.

\section{Privacy Preserving Data Imputation}
In our methods, except the datatype array indicating whether the feature is categorical or numeric, we keep every other information in a secret shared form. Along with the data matrix, we also keep an auxiliary matrix indicating missing features as $0$ and non-missing ones as $1$. Both of these matrices are secret shared.

\subsection{Privacy Preserving Mean Imputation}
The mean is calculated by summing all values and dividing by the count of non-missing values, obtained by summing the corresponding column in the auxiliary matrix. For mode calculation of categorical features in one-hot encoding format, we compare the mean to 0.5 using the \textit{Compare} function. Once mean or mode is obtained, an array indicating whether data needs imputation is created by subtracting a column of solely $1$'s from the corresponding column in the auxiliary matrix. This yields an array where $1$ indicates missing values and $0$ indicates non-missing values. This resulting vector is then multiplied by a vector of repeated mean (or mode), leaving mean only in rows requiring imputation, and $0$ otherwise. This is added to all values and assigned to the original data, resulting in mean-imputed data without missing values.

\subsection{Privacy Preserving Median Imputation}
Identifying the median in MPC is more complex than calculating the mean. Initially, the count of non-missing values is obtained by summing the respective feature column of the missing value auxiliary matrix. This count is divided by two using the \textit{Divide} function to find the index of the middle value. To address the odd number of samples with the non-missing feature, we round the resulting value by eliminating the fractional parts, representing the decimal part of the value. For an even count, one of the two middle values is selected, avoiding the need for averaging.

After obtaining the median index in secret share form, the target feature is sorted in descending order. Note that we set the missing features to the lowest possible value before sorting. This facilitates locating absent values at the lower end, crucial for median calculation. Two vectors, $\ell1$ and $\ell2$, are created, with $\ell1$ containing numbers from $1$ to the size of the feature vector and $\ell2$ consisting of the repeated median index. These vectors are privately compared using the \textit{Equals} function, producing a vector $C$ containing all $0$'s except for the position of the median. Vector $C$ is multiplied by the sorted feature vector, and the product is summed up, resulting in the median value. Before the final step, we set the missing features back to $0$. Finally, missing values are imputed using the median as we did in the mean imputation.

\subsection{Privacy Preserving Regression Imputation}
To implement regression imputation using MPC, we first implement multiple regression. This involves multiplying the data matrix $\mathcal{X}$ without the target feature by its transpose $\mathcal{X^T}$, followed by LU decomposition of the product $\mathcal{X^T X}$. LU decomposition results in lower triangular matrix $\mathcal{L}$ and upper triangular matrix $\mathcal{U}$. Then, $\mathcal{X^T}$ is multiplied by the target feature vector $y$. Forward substitution is performed to solve the linear system $\mathcal{L}z = \mathcal{X}^Ty$, followed by backward substitution to solve $\mathcal{U}\boldsymbol{\hat{\beta}} = z$ for the vector $\boldsymbol{\hat{\beta}}$, containing regression coefficients.

Regression imputation utilizes multiple linear regression with LU decomposition to predict missing values. Given an auxiliary matrix indicating missing values, the function takes the data matrix and the column number of the feature to be imputed. Rows with missing data are located and excluded, ensuring they do not impact regression. An intercept term is added by including a column of ones at the beginning. Rows with missing values are then excluded from the matrix. The column of the targeted feature is removed, and regression is performed using the modified matrix and the target feature array. The resulting coefficient vector $\boldsymbol{\hat{\beta}}$ is used to predict the target feature values regardless of whether they are missing.

To impute missing data with predicted values, the auxiliary matrix column corresponding to the target feature is compared to an array of zeros. Missing values are indicated by $1$, others by $0$. This result is multiplied by the predicted values and added to the target feature. Only missing values are imputed, leaving others unchanged. Finally, these changes are applied to the original matrix.

\subsection{Privacy Preserving kNN Imputation}
For kNN imputation in our framework, we utilize a function for kNN inference. Before performing kNN inference, the data matrix is modified so that features of rows containing missing values in the target column are set to $0$, achieved by element-wise multiplication with the auxiliary matrix. To predict a label for a new query, we calculate the squared Euclidean distance between the query and each sample in the dataset. Then, we identify the closest samples based on these distances without sorting the distances themselves, instead generating a sorting permutation based on distances.

The $k$-nearest neighbors' labels from the sorted labels array are chosen. Depending on the data type (numeric, binary, or multi-class), different strategies are employed. For numeric data, the average of the top $k$ labels is calculated. For binary data, the majority is determined by summing the labels of the $k$-nearest samples and comparing it with $k/2$. For multi-class data, a binary-like encoding method is used, and the majority class is identified based on the most significant bit of each label's binary encoding.

To impute with kNN inference results, an array indicating missing values is derived from the auxiliary matrix. This array, containing $1$'s where values are missing, is multiplied with the prediction results array, retaining only instances where data is absent. This result is added to the labels array, imputing missing data with predicted values while leaving non-missing ones unchanged. Finally, this imputation result is assigned to the original data matrix.

\section{Results}

\subsection{Dataset}
In our experiments, we utilize a diabetes dataset obtained from Kaggle\footnote{https://www.kaggle.com/datasets/iammustafatz/diabetes-prediction-dataset}, comprising $10,000$ records and $9$ attributes including gender, age, hypertension, heart disease, smoking history, BMI, HbA1c level, blood glucose level, and diabetes status. Categorical features like ``Gender'' and ``Smoking History'' are transformed into binary variables. In the latter case, ``current'' and ``former'' are encoded as $1$, while others are encoded as $0$. To assess correctness, we systematically delete every $10$th value from the target feature, introducing a $10\%$ missing data rate. For scalability analysis, varying sizes of dataset subsets are used to simulate different dataset magnitudes. This enables us to evaluate imputation methods' performance across datasets of different sizes, providing insights into their efficiency and scalability in data imputation.

\subsection{Experimental Setup}
In our experiments, we utilized Ubuntu operating system deployed within the Windows Subsystem for Linux (WSL) on a host machine running Windows 11. The Ubuntu distribution used was Ubuntu 20.04.6 LTS. The host machine employed for experimentation was equipped with an Intel(R) Core(TM) i7-1065G7 processor running at 2.6 GHz and 32 GB RAM. We set the number of bits to represent decimal points to $15$ in the number format unless otherwise stated.

\subsection{Correctness Analysis}
To verify the correctness of our proposed privacy-preserving data imputation methods on the diabetes dataset, we created datasets containing randomly selected $100$ and $1000$ samples, respectively. We repeat each experiment $10$ times and report the average of these runs.

\subsubsection{Privacy Preserving Mean Imputation} To check if the mean imputation for numerical values is correct, we fill in missing data for the BMI feature using our privacy-preserving mean imputation method. We also do the same for these datasets without privacy protection, using the standard mean filling. Then, we find the average absolute difference between all the values in the standard mean-filled dataset and the privacy-protected one. The results are shown in Table \ref{tab:non-cat_results}. To check if mean imputation works well for categorical values, we fill in missing data for the hypertension feature. Here, $1$ means the patient has hypertension. The results on the categorical features are displayed in Table \ref{tab:cat_results}.

\subsubsection{Privacy Preserving Median Imputation} 
To check if median imputation works correctly, we also use the BMI feature. We compare the result of our privacy-preserving median imputation method to the output of the plaintext and non-private median imputation method. We calculate the average absolute difference between the results of these methods on all imputed values. They indicate that our method produces the same result as the non-private version with a negligible error. The results of privacy-preserving median imputation of non-categorical features are shown in Table \ref{tab:non-cat_results}.

\begin{table}[!ht]
\centering
\begin{tabular}{ccc}
\hline
Method                                 & \multicolumn{1}{c}{Size} & \multicolumn{1}{c}{MAE} \\ \hline
\multirow{2}{*}{Mean Imputation}       & 100                      & $2.1 \times 10^{-5}$      \\ \cline{2-3} 
                                       & 1000                     & $1.8 \times 10^{-5}$      \\ \hline
\multirow{2}{*}{Median Imputation}     & 100                      & $1.9 \times 10^{-5}$      \\ \cline{2-3} 
                                       & 1000                     & $1.7 \times 10^{-5}$      \\ \hline
\multirow{2}{*}{Regression Imputation} & 100                      & $9.2 \times 10^{-6}$      \\ \cline{2-3} 
                                       & 1000                     & $2.1 \times 10^{-5}$      \\ \hline
\multirow{2}{*}{kNN Imputation}        & 100                      & $2.1 \times 10^{-5}$      \\ \cline{2-3} 
                                       & 1000                     & $1.8 \times 10^{-5}$      \\ \hline
\end{tabular}
\vspace{0.1cm}
\caption{The average mean absolute error (MAE) of $10$ repetitions of the correctness analysis on non-categorical features}
\label{tab:non-cat_results}
\end{table}

\begin{table}[!ht]
\centering
\begin{tabular}{ccc}
\hline
Method                                 & \multicolumn{1}{c}{Size} & \multicolumn{1}{c}{MAE} \\ \hline
\multirow{2}{*}{Mean Imputation} & 100 & $0.0$ \\ \cline{2-3} 
                                 & 1000 & $0.0$ \\ \hline
\multirow{2}{*}{kNN Imputation}  & 100 & $0.0$ \\ \cline{2-3} 
                                 & 1000 & $0.0$ \\ \hline
\end{tabular}
\vspace{0.1cm}
\caption{The average mean absolute error (MAE) of $10$ repetitions of the correctness analysis on categorical features}
\label{tab:cat_results}
\end{table}

\subsubsection{Privacy Preserving Regression Imputation}
To verify the accuracy of our privacy-preserving regression imputation, we impute the missing data for the blood glucose level feature, considering that there is a linear relation with other features. We compare this with a standard version of regression using LU decomposition without privacy protection. We again use the mean absolute error to measure the performance of our imputation method. These results are summarized in Table \ref{tab:non-cat_results}. Note that we set the number of bits representing decimal values to $18$ for the privacy-preserving regression imputation due to the high precision requirement.

\subsubsection{kNN Imputation}
To confirm if our privacy-preserving kNN imputation for numerical values is accurate, we focus on the imputation of missing data for the BMI feature. We compare this with sklearn's \textit{KNNImputer} package in Python. We set $k$ as $\sqrt{N}$, where $N$ is the number of samples \cite{Ispirova.2020}. So, for $N = 100$, $k = 10$, and for $N = 1000$, $k = 31$. Then, we find the average absolute difference between all values in the standard imputed dataset and the resulting dataset from our privacy-preserving kNN imputation method. The results of our method on non-categorical features are demonstrated in Table \ref{tab:non-cat_results}. We also analyzed the accuracy of our privacy-preserving kNN imputation method on categorical features. To achieve this goal, we select the hypertension feature. The results, which are given in Table \ref{tab:cat_results}, indicate that our method exactly realizes the non-private kNN imputation from sklearn.

\subsection{Performance Analysis}
In addition to the correctness analysis proving that we realize the non-private corresponding imputation methods in a privacy-preserving way, we also analyze the performance of our privacy-preserving imputation methods. Such an analysis is a significant component of our methodology, considering the importance of having scalable solutions to address real-life medical and healthcare problems. In our analysis, we vary the number of samples by doubling it in each step to see how well our solutions scale to varying numbers of samples. We use the same feature to impute for each imputation method as we did in the correctness analysis.

\subsubsection{Privacy Preserving Mean Imputation}
The results of our performance analysis experiments demonstrate that our method scales almost linearly to the number of samples in the dataset. Figure \ref{fig:performance_mean} summarizes this trend.

\subsubsection{Privacy Preserving Median Imputation}
The results of our scalability experiments indicate the near-linear execution time of our privacy-preserving median imputation method. This linear trend can be seen in Figure \ref{fig:performance_median}.

\subsubsection{Privacy Preserving Regression Imputation}
Our experiments using our method on varying numbers of samples show a very close resemblance to linear execution time. The Figure \ref{fig:performance_regression} demonstrates this trend.

\subsubsection{Privacy Preserving kNN Imputation}
\begin{figure}[!ht]
    \centering
    \begin{subfigure}[b]{0.408\linewidth}
        \centering
        \includegraphics[width=\linewidth]{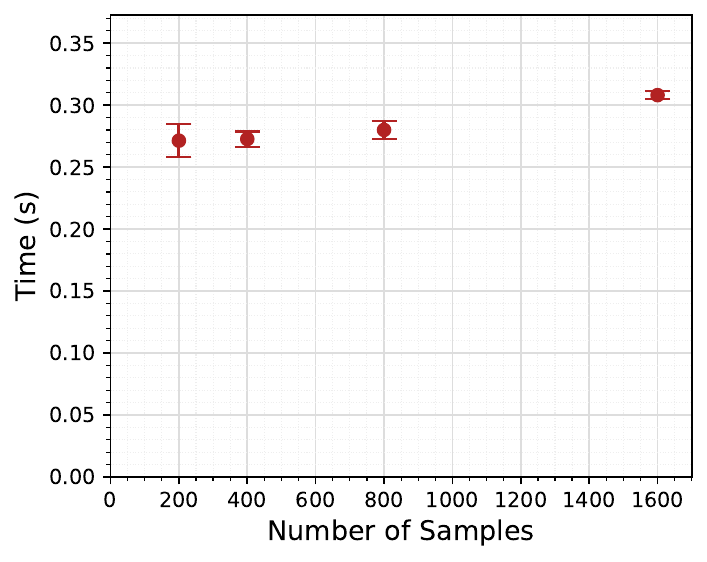}
        \caption{Mean Imputation}
        \label{fig:performance_mean}
    \end{subfigure}
    \begin{subfigure}[b]{0.4\linewidth}
        \centering
        \includegraphics[width=\linewidth]{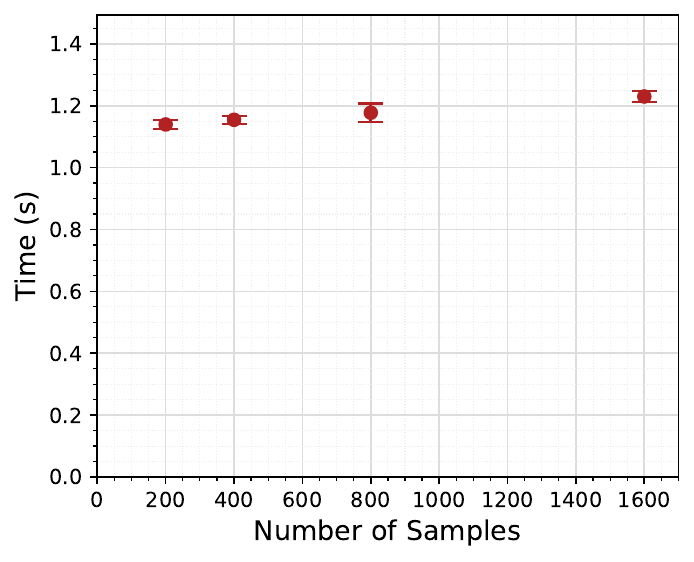}
        \caption{Median Imputation}
        \label{fig:performance_median}
    \end{subfigure}
    \vskip\baselineskip
    \begin{subfigure}[b]{0.4\linewidth}
        \centering
        \includegraphics[width=\linewidth]{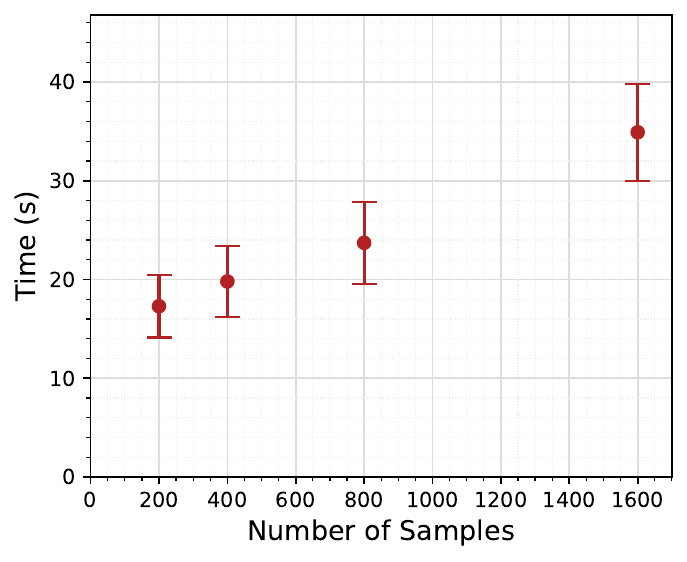}
        \caption{Regression Imputation}
        \label{fig:performance_regression}
    \end{subfigure}
    \begin{subfigure}[b]{0.43\linewidth}
        \centering
        \includegraphics[width=\linewidth]{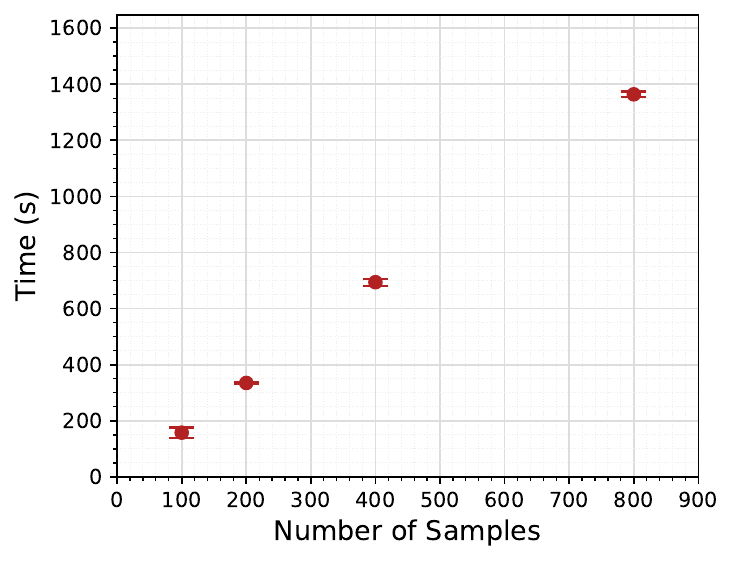}
        \caption{kNN Imputation}
        \label{fig:performance_knn}
    \end{subfigure}
    \caption{The performance analysis of our privacy-preserving data imputation methods on varying numbers of sample sizes}
    \label{fig:performance_all}
\end{figure}

We assessed our method's execution time on varying numbers of samples and Figure \ref{fig:performance_knn} summarizes these results. It indicates that our privacy-preserving kNN imputation scales linearly to the number of samples in the dataset. Even though there is a linear trend, the actual execution time of this method is significantly larger than other methods. The underlying reason for this behavior is the requirement of computation of all pairwise distances of the samples in the dataset. Since the computing parties are not aware of which samples are missing the corresponding feature, the computing parties compute each sample's distance to the rest of them, sort these distances, find the average or majority of the $k$-nearest neighbor, and assign this value to the feature of this sample if it is missing. Such an exhaustive operation makes our privacy-preserving kNN imputation relatively slower compared to other methods.


\section{Discussion}
Our experiments demonstrated that all our imputation methods are accurate, yielding results nearly identical to plaintext versions with minimal error. The largest absolute difference observed was $3 \times 10^{-3}$ or 0$.011\%$ of the value, and the overall mean absolute difference was around $2 \times 10^{-5}$. This minor error stems from the number format used in MPC, which has limited precision, and the multiplication function, where truncation errors can accumulate.


As expected, mean and median imputation, being computationally simple, are very efficient and can handle large datasets in seconds. For $1600$ samples, mean imputation takes $0.293$ seconds, and median imputation takes $1.261$ seconds. Although these methods are faster and preferable to listwise deletion, they may not account for relationships between features, potentially leading to biased estimates. Regression imputation, although slower, generally produces more accurate results by considering linear relationships between features. It takes $34$ seconds for $1600$ samples but remains computationally effective. There is room for performance improvement.

Using the building blocks, we successfully implemented kNN imputation. However, with almost $24$ minutes needed for $800$ samples, it is significantly slower. The main reason for such a long execution time is that the computing parties have to perform kNN imputation on every sample as if they would use the result of the imputation since they are unaware of the information of which samples are missing that feature due to privacy concerns. Considering that the required operations for kNN imputation in MPC are also time-consuming, the resulting execution time of kNN imputation is longer than the rest. While effective, kNN imputation is currently impractical for very large datasets in MPC.

In our methods, we use the semi-honest adversarial model as a security model, where the attacker follows the protocol but tries to deduce information about the data. Except for the datatype matrix, an attacker corrupting a single computing party cannot deduce the data, the auxiliary matrix, or any results at all. The overall security of our methods depends on the hybrid model, in which we use individually secure building blocks. For further details about the building blocks, please refer to \cite{Unal.07.02.2022}.

\section{Conclusion}
Compared to other steps of a machine learning application pipeline, the data preprocessing step has been overlooked in the literature. Even though the training, inference, and evaluation steps of a machine learning pipeline have been addressed in a privacy-preserving way, the data preprocessing step has not received well-deserved attention. Considering the fact that the success of the rest of the steps depends on a successful data preprocessing step, this urged us to focus on this problem. Therefore, in this paper, we addressed the data preprocessing step of machine learning applications in medical and healthcare domains by proposing privacy-preserving data imputation methods using MPC. We proposed mean, median, regression, and kNN imputation methods in a privacy-preserving way. Even though the mean and median imputation methods are relatively simple, this study fills a gap in the literature by implementing kNN and regression imputation within an MPC framework for the first time. It lays the foundation for organizations to perform collaborative data analysis with missing values while preserving privacy, which is particularly relevant in healthcare research where data sharing is necessary but privacy regulations are strict. These imputation methods enhance data quality for analysis and decision-making, allowing practitioners to use MPC for preprocessing data and improving overall results.

\bibliographystyle{unsrt}  
\bibliography{ref}

\end{document}